\documentclass[floatfix,aps,showpacs,twocolumn,prb]{revtex4}%
\usepackage{amsmath}
\usepackage{amssymb}
\usepackage{graphicx}
\usepackage{dcolumn}
\usepackage{natbib}
\usepackage{bm}
\usepackage{epsfig}
\usepackage{amsfonts}
\usepackage{color}%
\providecommand{\U}[1]{\protect\rule{.1in}{.1in}}
\providecommand{\U}[1]{\protect\rule{.1in}{.1in}}
\providecommand{\U}[1]{\protect\rule{.1in}{.1in}}
\providecommand{\U}[1]{\protect\rule{.1in}{.1in}}
\providecommand{\U}[1]{\protect\rule{.1in}{.1in}}
\providecommand{\U}[1]{\protect\rule{.1in}{.1in}}

\begin{document}
\title{Disorder effect of resonant spin Hall effect in a tilted magnetic field}
\author{Zhan-Feng Jiang, Shun-Qing Shen and Fu-Chun Zhang}
\affiliation{Department of Physics, and Center of Theoretical and Computational Physics,
The University of Hong Kong, Pokfulam Road, Hong Kong}
\date{\today }

\begin{abstract}
We study the disorder effect of resonant spin Hall effect in a two-dimension
electron system with Rashba coupling in the presence of a tilted magnetic
field. The competition between the Rashba coupling and the Zeeman coupling
leads to the energy crossing of the Landau levels, which gives rise to the
resonant spin Hall effect. Utilizing the Streda's formula within the
self-consistent Born approximation, we find that the impurity scattering
broadens the energy levels, and the resonant spin Hall conductance exhibits a
double peak around the resonant point, which is recovered in an applied titled
magnetic field.

\end{abstract}

\maketitle

\section{Introduction}

Spin-orbit couplings open a route to control quantum electron spin by electric
means. One of the efficient methods to inject or generate electron spin in
non-magnetic semiconductors is the spin Hall effect, in which an electric
current or an electric field may induce a transverse spin current in the
systems with strong spin-orbit couplings. Early theories
\cite{Dyakonov71,Hirsch99} proposed that the spin current is caused by
asymmetric scattering of electrons with spin-up and -down in impurity
potentials, named as extrinsic spin Hall effect. In recent years it was
demonstrated that the spin-orbit coupling in the electron bands can also lead
to an intrinsic spin Hall effect in either p-doped or n-doped
semiconductors.\cite{Zhangshoucheng,Niuqian} Both extrinsic and intrinsic spin
Hall effects were confirmed experimentally in various
systems.\cite{Kato04Science,Wunderlich05PRL,Valenzuela06Nature,Seki08NM}

A two-dimensional electron gas (2DEG) with a Rashba coupling was proposed to
exhibit an intrinsic spin Hall effect.\cite{Niuqian,Shen-04prb} The spin-orbit
coupling in 2DEG modifies the electron band structure, and may lead to
interesting magnetotransport properties, such as the beating phenomenon in the
Shubnikov-de Haas (SdH) oscillation.\cite{SdH,parameters} When the system is
subjected to an external magnetic field, the Zeeman splitting will also change
the spin-dependent electron bands. The interplay of the spin-orbit coupling
and the Zeeman coupling produces the crossing of electron energy levels. Based
on this property, it was proposed that a tiny electric field may remove the
additional degeneracy of energy levels and produces a finite spin current if
the Fermi surface sweeps across the crossing point of energy levels. As a
result, there exhibits a divergent spin Hall
conductance.\cite{Shen,Shen2,Bao-05PRB,Zhang-08IJMPB} This resonant spin Hall
effect was also discussed in p-doped systems in a magnetic
field.\cite{Matianxing,Zarea06PRB}

However, impurities in the system make the issue more
subtle.\cite{Sinova-SSC,Schliemann} The vertex correction in the self energy
turns out to cancel the spin Hall conductance even in a weak disorder limit in
2DEG with linear Rashba coupling, while the spin Hall conductance survives in
p-doped Luttinger model and the systems with cubic spin-orbit
couplings.\cite{SHC-Inoue,SHC-Murakami,SHC-Dimitrova,SHC-Raimondi,SHC-Grimaldi,Liuchaoxing}
The disorder effect strongly depends on the symmetry of the spin-orbital
coupling and the dispersion. Now whether the resonant spin Hall effect can
survive in a finite density of impurities becomes an issue to be answered.
This is the motivation of the present work.

Here we present a full investigation on the disorder effect of resonant spin
Hall effect in 2DEG with the Rashba coupling in a tilted magnetic field. The
impurity effect is considered by the self-consistent Born approximation (SCBA)
and the vertex correction in the ladder approximation. We found the impurity
effect will suppress the resonant spin Hall conductance at the resonant point,
and produce a double peak structure of the spin Hall conductance around the
point. A tilted magnetic field is applied to enhance the effective Zeeman
splitting,\cite{SdH2,Du-09exp,Bychkov,Das,Wilde} and to recover the effect
when the energy level splitting is larger than the energy broadening by impurities

\section{General formalism}

\label{sec_formulism}

\subsection{2DEG in a tilted field}

\label{sec_Hamiltonian}

We consider a 2DEG in the x-y plane with the Rashba spin-orbit interaction in
a tilted magnetic field. The perpendicular component of the tilted field is
$-B_{\bot},$ and the in-plane component is chosen to be along the x-direction
$B_{\bot}\tan\theta$, where $\theta$ is the angle between the field and the
z-direction. We take the Laudau gauge for the vector potential of the field
$\overrightarrow{B}=(B_{\bot}\tan\theta,0,-B_{\bot})$. The total Hamiltonian
including the Zeeman energy is given by%
\begin{align}
H_{0}  &  =\frac{1}{2m}[(p_{x}+eB_{\bot}y)^{2}+p_{y}^{2}]+\frac{\lambda}%
{\hbar}[(p_{x}+eB_{\bot}y)\sigma_{y}-p_{y}\sigma_{x}]\nonumber\\
&  -\frac{1}{2}g_{s}\mu_{B}B_{\bot}\sigma_{z}+\frac{1}{2}g_{s}\mu_{B}B_{\bot
}\tan\theta\sigma_{x}, \label{H_SOC}%
\end{align}
where $\mathbf{p}=-i\hbar\bigtriangledown,$ $m,-e,g_{s}$ are the electron's
effective mass, charge, and Lande $g$ factor, respectively. $\mu_{B}$ is the
Bohr magneton, $\lambda$ is the strength of Rashba spin-orbit coupling, and
$\sigma_{i}$ are the Pauli matrices. We take a periodic boundary condition
along the x direction, hence the momentum $p_{x}=\hbar k$ is a good quantum number.

An analytical solution can be obtained in the case of $\theta=0$%
.\cite{Rashba,Schliemann03PRB,Shen} There were some studies on spin transport
based on the solution.\cite{Shen2,Bao,Lucignano08PRB} Inclusion of the tilted
field makes the problem much more complicated, and an analytical solution is
not available at present. In the following approach, we choose the energy
eigenstates for the system without the spin-orbit coupling ($\lambda=0$) and
$\theta=0$ as a set of basis,
\begin{equation}
\left\vert nk\sigma\right\rangle =\frac{1}{\sqrt{L_{x}}}e^{ikx}\phi
_{n}(y+kl_{b}^{2})\left\vert \sigma\right\rangle , \label{basis}%
\end{equation}
where the magnetic length $l_{b}=\sqrt{\hbar/eB_{\bot}}$, the spin index
$\sigma=\uparrow,\downarrow$ and $L_{x(y)}$ is the length of the 2DEG,
$\phi_{n}(y)$ is the eigenstate of the $n^{th}$ energy level of a linear
oscillator with the frequency $\omega=eB_{\perp}/m$,\cite{Landau} and
$\left\vert \sigma\right\rangle $ is the eigenstate of spin $\sigma_{z}$.

When the tilt angle $\theta=0$, the system can be solved exactly.\cite{Rashba,
Shen} The eigenvalues of $H_{0}$ are given by%
\begin{equation}
\epsilon_{ns}=\hbar\omega(n+\frac{s}{2}\sqrt{(1-g)^{2}+8n\eta^{2}}),
\end{equation}
where $\eta=\lambda ml_{b}/\hbar^{2}$ and $g=g_{s}m/2m_{e}$ with $m_{e}$ the
mass of a free electron, $s=1$ for $n=0$, and $s=\pm1$ for $n\geqslant1$. The
states $\Phi_{nks}$ have a degeneracy $N_{\phi}=L_{x}L_{y}eB/h$, corresponding
to $N_{\phi}$ values of $k$. The eigenstate has the form,%
\begin{equation}
\Phi_{nks}=\cos\theta_{ns}\left\vert n,k,\uparrow\right\rangle +i\sin
\theta_{ns}\left\vert n-1,k,\downarrow\right\rangle ,
\end{equation}
where $\theta_{01}=0$, and for $n\geqslant1$, $\theta_{ns}=\arctan
(-u_{n}+s\sqrt{1+u_{n}^{2}})$ with $u_{n}=(1-g)/\sqrt{8n}\eta$. One of the
features of the solution is the crossing of the energy levels as functions of
the magnetic field, which is caused by the competition between the spin-orbit
coupling and Zeeman energy splitting. For the two levels $\epsilon_{n1}$ and
$\epsilon_{n+1,-1}$, the condition for the crossing is determined
by\cite{Shen}%
\begin{equation}
\sqrt{(1-g)^{2}+8n\eta^{2}}+\sqrt{(1-g)^{2}+8(n+1)\eta^{2}}=2.
\end{equation}
This point is called the resonant point for resonant spin Hall effect.

This additional degeneracy due to the competition between the spin-orbit
coupling and the Zeeman energy of the perpendicular field can be removed by a
tilted field. In the case of $\theta\neq0,$ the energy levels can be
calculated numerically. Using the expression in Eq. (4), we may make a
truncation approximation by keeping the Landau levels with $n<N$ such that the
dimensionality of matrix is reduced to $2N\times2N.$ Numerical diagonalization
of the matrix can give us the energy eigenvalues.

\begin{figure}[ptb]
\centering \includegraphics[width=0.45\textwidth]{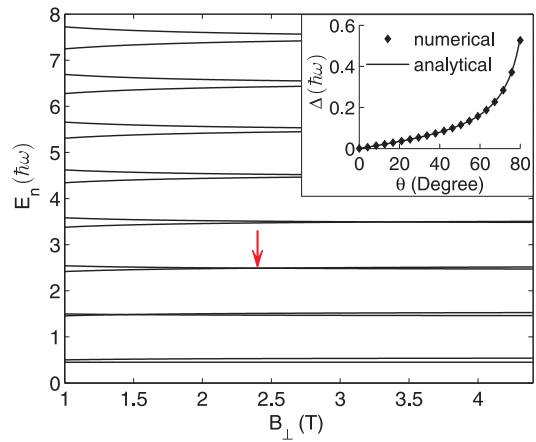}\caption{(Color
online) Energy levels as functions of the magnetic field when the tilt angle
$\theta=0$, the arrow denotes a level-crossing point, which develops into a
gap when the tilt angle increases. The inset shows the gap as a function of
the tilt angle, reflects the accordance of the numerical calculation and the
analytic expression Eq. (\ref{Delta}). The energy has been scaled by
$\hbar\omega=\hbar eB/m.$ }%
\label{gap}%
\end{figure}

Alternatively, the gap can also be calculated approximately by the degenerate
perturbation theory. We take the partial Hamiltonian $H^{\prime}=g_{s}\mu
_{B}B_{\bot}\tan\theta\sigma_{x}/2$ as a perturbation, and express it in the
subspace spanned by the two states $\Phi_{n,k,1}$ and $\Phi_{n+1,k,-1}$ near
the resonant point,%
\begin{equation}
\widetilde{H^{\prime}}=\left[
\begin{array}
[c]{cc}%
0 & i\Delta/2\\
-i\Delta/2 & 0
\end{array}
\right]  , \label{H'}%
\end{equation}
where the gap $\Delta$ is%
\begin{equation}
\Delta=g_{s}\mu_{B}B_{\bot}\tan\theta\cos\theta_{n1}\sin\theta_{n+1,-1}.
\label{Delta}%
\end{equation}
In Fig. \ref{gap}, we present the energy levels of $\theta=0$ as a function of
$B_{\bot}.$ The parameters used are $\lambda=9\times10^{-12}eVm$, $g_{s}=4$,
and $m=0.05m_{e}$.\cite{parameters} The red arrow denotes a level crossing at
$B_{0}\approx2.4T$. The insert shows the energy gap as a function of the tilt
angle $\theta$ with $B\bot=B_{0}$. We notice that the numerical and analytical
results are in good agreements.

\subsection{Self-Consistent Born Approximation}

\label{sec_SCBA}

In this section, we briefly review the general formalism of linear response
theory of self-consistent Born approximation (SCBA) for electron transport. We
shall use this technique to investigate the transport properties of 2DEG with
a Rashba coupling in a tilted magnetic field, especially near the resonant
point. The effect of impurities will be taken into account in this formalism.

We consider a random configuration of impurities with short-range potentials
$V(\mathbf{r})=\sum_{j=1}^{N_{i}}V\delta(\mathbf{r-R}_{j})$, where
$\mathbf{R}_{j}$ is the position of the $j^{th}$ impurity. The density of the
impurities is $n_{i}=N_{i}/(L_{x}L_{y})$. Generally speaking, the Green's
functions for a specific configuration\ of random potential can be written as
$\mathcal{G}^{\pm}(E)=[E-H_{0}-V(\mathbf{r})\pm i0^{+}]^{-1}$, where $+$ and
$-$ correspond to the retarded and advanced Green's function, respectively.
All transport quantities can be expressed in terms of the Green's function
after averaging all possible configurations of the impurities. Using the
conventional perturbation expansion with respect to $V(\mathbf{r})$, we can
obtain the Dyson equation for the averaged Green's function $G^{\pm}$. The
impurity effect is absorbed by a self-energy function $\Sigma^{\pm}$ as
follows,
\begin{equation}
G^{\pm}(E)\equiv\left\langle \mathcal{G}^{\pm}(E)\right\rangle _{c}%
=[E-H_{0}-\Sigma^{\pm}(E)],
\end{equation}
where $\left\langle \cdots\right\rangle _{c}$ means the average over all the
impurity configurations. In the SCBA, the self-energy operator can be
expressed by $\Sigma^{\pm}=\left\langle VG^{\pm}(E)V\right\rangle _{c}%
$.\cite{Gerhardts,Vasko,Rammer,Sadovskii} In the representation of the Landau
levels, $G$, $\Sigma$, and $V$ are expressed as matrices. For such a
spin-independent impurity potential, previous works\cite{Bastin,Gerhardts}
proved that the self-energy is independent of $n$ and $k$ for a
spin-independent Landau system. We find that the self-energies for a
spin-orbit coupling system are independent of $n$ and $k$,
\begin{equation}
\Sigma_{nk\sigma,n^{\prime}k^{\prime}\sigma^{\prime}}^{\pm}=\delta
_{nn^{\prime}}\delta_{kk^{\prime}}n_{i}V^{2}\frac{N_{\phi}}{L_{x}L_{y}%
}\underset{n_{1}}{%
{\displaystyle\sum}
}G_{n_{1}\sigma,n_{1}\sigma^{\prime}}^{\pm}. \label{self}%
\end{equation}
Here we dropped the index $k$ in $G^{\pm}$ because the averaged Green's
functions are k-independent.

\subsection{Streda's Formula for spin Hall conductivity}

With the averaged Green's function in mind, we can use the Kubo formula to
calculate the linear response of any physical quantity $\widehat{O}$ to an
external electric field $E_{ext}$ in the $\nu$ direction,
\begin{equation}
\sigma_{\nu}^{O}=\lim_{E_{ext}\rightarrow0}\left\langle \widehat
{O}\right\rangle _{c}/E_{ext}.
\end{equation}
As the single-particle version of the Kubo formula, the Streda's formula is a
conventional and powerful tool to study the transport property of 2DEG system
under a magnetic field.\cite{Streda} At the zero temperature, the formula is
given by%
\begin{align}
\sigma_{\nu}^{O}(E_{f})  &  =ie\hbar\left\langle \int_{-\infty}^{Ef}dE\cdot
Tr[\widehat{O}\frac{d\mathcal{G}^{+}(E)}{dE}v_{\nu}A(E)\right. \nonumber\\
&  \left.  -\widehat{O}A(E)v_{\nu}\frac{d\mathcal{G}^{-}(E)}{dE}]\right\rangle
_{c}, \label{O}%
\end{align}
where $A\equiv(\mathcal{G}^{-}-\mathcal{G}^{+})/(2\pi i)$ is the spectral
function and $v_{\nu}\equiv\frac{1}{i\hbar}[r_{\nu},H]$ is the velocity
operator,. Because there are products of two Green's functions in the impurity
average $\left\langle \cdots\right\rangle _{c}$, the vertex correction has to
be included. For a specific density of charge carriers, the Fermi energy as a
function of the magnetic field is determined by
\[
n_{e}=\left\langle \int_{-\infty}^{Ef}dE\cdot Tr[A]\right\rangle _{c}.
\]

Usually the Streda's formula is applied to calculate the electric conductance
by replacing $\widehat{O}$ by an electric current operator, $J_{\nu}=-ev_{\nu
}$. In the present work, we intend to explore the spin transport in the
system. The spin current is defined as $j_{\mu}^{\alpha}=(\hbar/4)\{v_{\mu
},\sigma_{\alpha}\}$, which is a tensor determined by both the motion
direction of an electron and its polarization. In the framework of linear
response theory, the spin Hall conductivity $\sigma_{\mu\nu}^{\alpha}$, the
ratio of the spin Hall current to an external field, can be calculated by
substituting $\hat{O}=j_{\mu}^{\alpha}$ ($\mu\neq\nu$) in Eq. (\ref{O}). The
spin Hall conductivity comes from the contribution of all the electrons below
the Fermi level. Opposite to those for the conductivity and Hall conductivity,
it cannot be reduced to a Fermi edge quantity\cite{Gerhardts} because the spin
current is not a commutator of any operator and the
Hamiltonian.\cite{Shijunren} For the purpose of our numerical calculation, we
transform the Streda's formula into the following form,%
\begin{equation}
\sigma_{\mu\nu}^{\alpha}(E_{f})=\frac{e\hbar}{2\pi}\int_{-\infty}^{Ef}dE\cdot
Tr[j_{\mu}^{\alpha}(K_{\nu}^{+-}-K_{\nu}^{++}-K_{\nu}^{--})],
\label{sigma_alpha_munu}%
\end{equation}
where
\begin{align}
K_{\nu}^{+-}  &  \equiv\frac{d\left\langle \mathcal{G}^{+}v_{\nu}%
\mathcal{G}^{-}\right\rangle _{c}}{dE},\label{K1}\\
K_{\nu}^{++}  &  \equiv\left\langle \frac{d\mathcal{G}^{+}}{dE}v_{\nu
}\mathcal{G}^{+}\right\rangle _{c},K_{\nu}^{--}=\left[  K_{\nu}^{++}\right]
^{+}. \label{K2}%
\end{align}
$K_{\nu}^{\sigma\sigma^{\prime}}$ are determined in a set of
Bethe-Salpeter-like equations,%

\begin{align}
K_{\nu}^{+-}  &  =\frac{dG^{+}}{dE}\left[  G^{+}\right]  ^{-1}F_{\nu}%
^{+-}+F_{\nu}^{+-}\left[  G^{-}\right]  ^{-1}\frac{dG^{-}}{dE}\nonumber\\
&  +G^{+}\left\langle VK_{\nu}^{+-}V\right\rangle _{c}G^{-},\label{K+-}\\
K_{\nu}^{++}  &  =\frac{dG^{+}}{dE}\left[  G^{+}\right]  ^{-1}F_{\nu}%
^{++}+G^{+}\left\langle VK_{\nu}^{++}V\right\rangle _{c}G^{+}.
\end{align}
where $F_{\nu}^{\sigma\sigma^{\prime}}=\left\langle \mathcal{G}^{\sigma}%
v_{\nu}\mathcal{G}^{\sigma^{\prime}}\right\rangle _{c}$ ($\sigma
,\sigma^{\prime}\in\{+,-\}$) is the vertex operator, which satisfy the
Bethe-Salpeter equation in the ladder approximation,%
\begin{equation}
F_{\nu}^{\sigma\sigma^{\prime}}=G^{\sigma}(v_{\nu}+\left\langle VF_{\nu
}^{\sigma\sigma^{\prime}}V\right\rangle _{c})G^{\sigma^{\prime}}. \label{BS}%
\end{equation}
These equations can be solved self-consistently and Eq. (\ref{K+-}) and
(\ref{BS}) have multi-solutions. From the continuity equation for charge
current in equilibrium, we can derive an auxiliary equation,%
\begin{equation}
Tr[F_{\nu}^{+-}\left(  \Sigma^{-}-\Sigma^{+}\right)  ]=0. \label{con1}%
\end{equation}
Differentiating Eq. (\ref{con1}) with respect to $E$ leads to another
auxiliary equation,%
\begin{equation}
Tr[K_{\nu}^{+-}\left(  \Sigma^{-}-\Sigma^{+}\right)  ]=-Tr[F_{\nu}^{+-}%
\frac{d\left(  \Sigma^{-}-\Sigma^{+}\right)  }{dE}].
\end{equation}
$dG^{\sigma}/dE$ is determined by an another self-consistent equation,
\begin{equation}
\frac{dG^{\sigma}}{dE}=-G^{\sigma}G^{\sigma}+G^{\sigma}\left\langle
V\frac{dG^{\sigma}}{dE}V\right\rangle _{c}G^{\sigma}.
\end{equation}
Since the density of states for each Landau level in SCBA has a semi-elliptic
form,\cite{Ando,Ando2,Girvin} it approaches zero and $dG^{\sigma}/dE$ becomes
infinity at the edge of each level. We find the integrand in Eq.
(\ref{sigma_alpha_munu}) is always convergent in numerical calculation,
because the concurrence of $d\mathcal{G}^{\sigma}/dE$ and $\mathcal{G}%
^{\sigma}$ in Eq. (\ref{K1}), (\ref{K2}).

\section{Numerical Results}

\label{sec_results}

Now we are ready to calculate the spin Hall conductivities numerically. In
this paper, the electron density is fixed at $n_{e}=2.9\times10^{15}m^{-2}$.
This value of $n_{e}$ promises that the Fermi level is located near the
resonant point with the filling factor $\nu=n_{e}/(N_{\phi}/L_{x}L_{y})=5$
while the magnetic field sweeps over the point, i.e., $B_{\perp}=B_{0}$ as
indicated in Fig. \ref{gap}. The other parameters are as the same as those
used in Fig. \ref{gap}. The density of impurities $n_{i}$ and the strength of
the impurity potential $V$ are combined in one parameter \textquotedblleft
scattering strength\textquotedblright\ $\Gamma\equiv n_{i}V^{2}m/(2\pi
\hbar^{2})$. We assign $\Gamma$ various values to investigate the impurity effect.

\subsection{Disorder effect of Resonant spin Hall effect}

\begin{figure}[ptb]
\centering \includegraphics[width=0.45\textwidth]{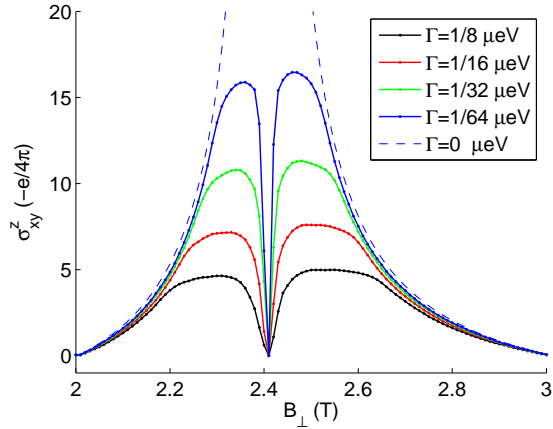}\caption{(Color
online) The spin Hall conductivity as a function of the magnetic field around
the level-crossing point when $\theta=0$, for various scattering strength
$\Gamma$. }%
\label{sigmazxy_impurity}%
\end{figure}

We first discuss the disorder effect of spin Hall conductance, especially near
the crossing point. We apply the formula in Eq.(\ref{sigma_alpha_munu}) to
calculate the spin Hall conductivity $\sigma_{xy}^{z}$ around the resonant
point $B_{0}=2.4T$ numerically for various strengths of disorder. Numerical
results are plotted in Fig. \ref{sigmazxy_impurity}. The dashed curve for
$\Gamma=0$ is from the solution in Ref. [\onlinecite{Shen}]. The key feature
of the disorder effect is the suppression of the spin Hall conductivity at the
resonant poin\.{t}. The large spin Hall conductance exhibits when the field
deviates from the resonant point and forms a double peak structure. The weight
of the spin Hall conductivity increases as the impurity strength decreases,
which reflects the intrinsic properties of the resonance. \begin{figure}[ptb]
\centering
\includegraphics[trim=0.3in 0.000000in 0.3in 0.6in, width=0.45\textwidth]{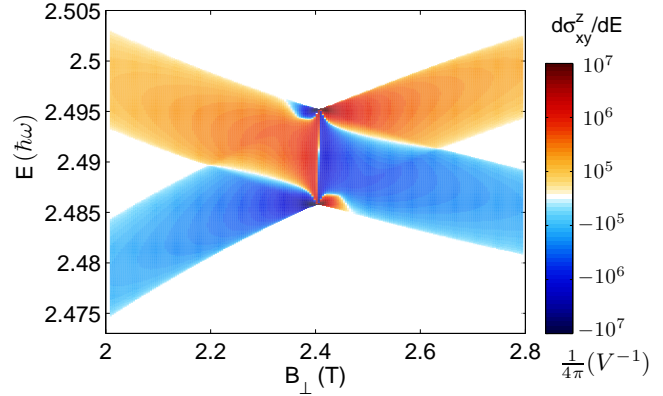}\caption{(Color
online) The distribution of the spin\ Hall conductivity on the electron's
energy, $d\sigma_{xy}^{z}/dE$, as a function of the energy $E$ and the
magnetic field $B_{\bot}$ when the tilt angle $\theta=0$, with the impurity
strength $\Gamma=1/32\mu eV$. The energy $E$ has been scaled by $\hbar\omega.$
}%
\label{cross}%
\end{figure}

To understand the suppression of resonant spin Hall conductance, we plotted in
Fig. \ref{cross} the distribution of the spin Hall conductivity on the
electron's energy, $d\sigma_{xy}^{z}/dE$, around the crossing point. The total
spin Hall conductivity $\sigma_{xy}^{z}(E_{f})=\int_{-\infty}^{E_{f}}%
(d\sigma_{xy}^{z}/dE)dE$ are contributed by all the electron states under the
Fermi level. It was observed that the energy levels are broadened due to the
impurity scattering, and the distribution of the spin Hall conductivity is
inhomogeneous. So the magnitude and even the sign of the spin Hall
conductivity can be varied by the Fermi level or the electron density. In the
clean limit, a tiny external electric field can open a gap between the
crossing levels, which leads to spin Hall conductance divergent. However,
after the impurity scattering is taken into account, a tiny external field
cannot open an energy gap any more because of the level broadening. As a
result, the spin Hall conductance in a weak field limit will be suppressed.
However, once the external field becomes stronger than the level broadening, a
large spin Hall conductance will appear. This can be seen from the case that
the magnetic field deviates from the crossing point, i.e., the additional
degeneracy of the two levels will be lifted, and a strong spin Hall
conductance recovers. This is the physical origin of the double peak structure
of the resonant spin Hall conductance. It is worth stessing that this
suppression of resonant spin Hall conductance is different from the case in
the absence of the Zeeman term. The Zeeman splitting may produce a non-zero
spin Hall conductance in the Rashba system.\cite{Bao-05PRB}

\subsection{Effect of a tilted field}

\begin{figure}[ptb]
\centering \includegraphics[width=0.45\textwidth]{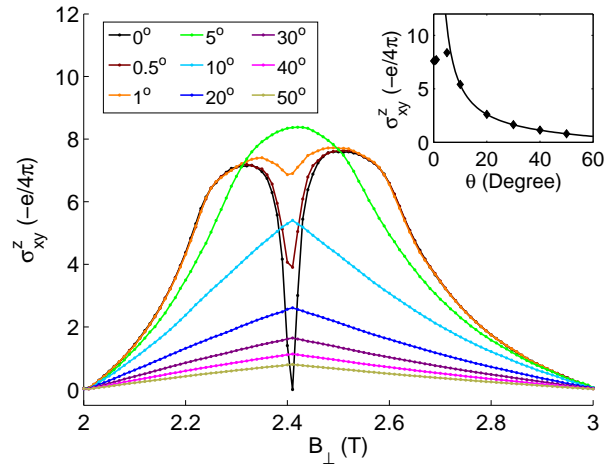}\caption{(Color
online) Spin Hall conductivity as a function of the perpendicular component of
the magnetic field for various tilt angles $\theta$. The insert compares the
peak height of the spin Hall conductivity as function of $\theta$ from the
numerical calculation (diamond) and from the analytical formula Eq.
(\ref{sigmazxy_analytic}) (solid line). The scattering strength $\Gamma
=1/16\mu eV$. }%
\label{sigmazxy_theta}%
\end{figure}

To further illustrate the formation of the resonant spin Hall effect, we
investigate the effect of the tilted magnetic field. Fig. \ref{sigmazxy_theta}
shows the dependence of the spin Hall conductivity on the tilt angle near the
resonant point. For the purpose of numerical calculation, we take the
scattering strength $\Gamma=1/16\mu eV$. As the tilted angle increases, the
spin Hall conductivity at the resonant point will increase very quickly, and
the two peaks finally integrate into one. After that point, the spin Hall
conductivity begins to decrease. These behaviors can be understood as the
competition between the disorder broadening of the energy levels and the
degeneracy lifting by the tilted field.

In the clean limit, the tilted field will remove the degeneracy of the energy
crossing levels as shown in Fig. \ref{gap}. We can estimate the peak height of
the spin Hall conductivity as a function of the the tilt angle by a
perturbation calculation adopted in Sec. \ref{sec_Hamiltonian}. Diagonalizing
the truncated two-level Hamiltonian in Eq. (\ref{H'}), we get the modified
eigenstates $\Psi_{\pm}=\left(  \Phi_{nk1}\pm i\Phi_{n+1,k,-1}\right)
/\sqrt{2}$, and the energy correction $E_{\pm}=\pm\Delta/2$ with Eq.
(\ref{Delta}). If the Fermi level just lies between the energy levels, the
spin Hall conductivity is mainly attributed to $\Psi_{-}$, which can be
calculated by the Kubo formula,\cite{Shen, Shen2}
\begin{align}
\sigma_{xy}^{z(1)}(B_{0})  &  =\underset{k}{%
{\displaystyle\sum}
}(\frac{\left\langle \Psi_{+}\right\vert ey\left\vert \Psi_{-}\right\rangle
\left\langle \Psi_{-}\right\vert j_{x}^{z}\left\vert \Psi_{+}\right\rangle
}{E_{-}-E_{+}}+c.c)\nonumber\\
&  =\frac{\hbar e^{2}B_{0}}{4\pi m\Delta}[(n+1)\cos^{2}\theta_{n1}\cos
^{2}\theta_{n+1,-1}\nonumber\\
&  -n\sin^{2}\theta_{n1}\sin^{2}\theta_{n+1,-1}], \label{sigmazxy_analytic}%
\end{align}
where $y=(mv_{x}|_{\lambda=0}-\hbar k)/eB$.

The impurity scattering will cause the level broadening. If the level
broadening is larger than the gap caused by the tilted field, the impurity
effect is dominant. Otherwise the tilted field effect will be dominant. The
level broadening is characterized by the half-width of the semi-elliptic
density of states of the Landau levels, which can be estimated approximately
by\cite{Girvin} $\gamma=2\sqrt{n_{i}V^{2}N_{\Phi}/L_{x}L_{y}}$. When
$\Delta>2\gamma$, the effect of the tilted field turns out to be dominant. The
insert of Fig. \ref{sigmazxy_theta} presents the data of the peak height of
the spin Hall conductivity as a function of $\theta$ from the numerical
calculation (diamond) and from the analytic formula (solid line) in Eq.
(\ref{sigmazxy_analytic}), They are in a good agreement when $\theta
\geqslant10^{\circ}$, which is very close to the estimated value
$\theta\approx8^{\circ}$ from $\Delta=2\gamma$.

\section{Summary}

\label{sec_summary}

In summary, we applied the Streda's formula to study the disorder effect of
the resonant spin Hall effect in the 2DEG system with the Rashba interaction
in a tilted magnetic field. Considering the vertex corrections in the self
energy, we find that the main effect of the impurity scattering is to broaden
the Landau levels. In the framework of linear response, the electric field is
taken to approach zero, and the energy splitting caused by the electric field
is always less than the broadening of the Landau levels. Thus a tiny external
field cannot remove the additional degeneracy of the energy levels at the
resonant point. As a result, the spin Hall conductance will be suppressed at
the point. When the magnetic field slightly deviates the resonant point or a
tilted field is applied, the degeneracy will be removed, a large spin Hall
conductance will be recovered. The spin Hall conductance exhibits a double
peak around the resonant point. From the effect of a tilted field, we believe
that a finite electric field, if it is strong enough to overcome the energy
level broadening, will recover the spin Hall effect even at the resonant
point. This is quite different from the disorder effect of spin Hall effect in
the Rashba system in the absence of magnetic field.\cite{Sinova-SSC}

\begin{acknowledgments}
The work is partly motivated with the tilted magnetic field experiments in R.
R. Du's group, to whom we wish to thank for stimulating discussions. This work
was supported by the Research Grant Council of Hong Kong under Grant No.: HKU
7041/07P and HKU 10/CRF/08.
\end{acknowledgments}


\begin{thebibliography}{99}                                                                                               %


\bibitem {Dyakonov71}M. I. D'yakonov and V. I. Perel', JETP Lett. \textbf{13},
467 (1971); Phys. Lett. A \textbf{35}, 459 (1971).

\bibitem {Hirsch99}J. E. Hirsch, Phys. Rev. Lett. \textbf{83}, 1834 (1999).

\bibitem {Zhangshoucheng}S. Murakami, N. Nagaosa, and S. C. Zhang, Science
\textbf{301}, 1348 (2003).

\bibitem {Niuqian}J. Sinova, D. Culcer, Q. Niu, N. A. Sinitsyn, T. Jungwirth,
and A. H. MacDonald, Phys. Rev. Lett. \textbf{92}, 126603 (2004).

\bibitem {Kato04Science}Y. K. Kato, R. C. Myers, A. C. Gossard, and D. D.
Awschalom, Science \textbf{306}, 1910 (2004).

\bibitem {Wunderlich05PRL}J. Wunderlich, B. Kaestner, J. Sinova, and T.
Jungwirth, Phys. Rev. Lett. \textbf{94}, 047204 (2005).

\bibitem {Valenzuela06Nature}S. O. Valenzuela and M. Tinkham, Nature (London)
\textbf{442}, 176 (2006).

\bibitem {Seki08NM}T. Seki, Y. Hasegawa, S. Mitani, S. Takahashi, H. Imamura,
S. Maekawa, J. Nitta, and K. Takanashi, Nat. Mater. \textbf{7}, 125 (2008).

\bibitem {Shen-04prb}S. Q. Shen, Phys. Rev. B \textbf{70}, 081311(R) (2004).

\bibitem {SdH}G. Engels, J. Lange, Th. Sch\"{a}pers, and H. L\"{u}th, Phys.
Rev. B \textbf{55}, R1958 (1997).

\bibitem {parameters}J. Nitta, T. Akazaki, H. Takayanagi, and T. Enoki, Phys.
Rev. Lett. \textbf{78}, 1335 (1997).

\bibitem {Shen}S. Q. Shen, M. Ma, X. C. Xie, and F. C. Zhang, Phys. Rev. Lett.
\textbf{92}, 256603 (2004).

\bibitem {Shen2}S. Q. Shen, Y. J. Bao, M. Ma, X. C. Xie, and F. C. Zhang,
Phys. Rev. B \textbf{71}, 155316 (2005).

\bibitem {Bao-05PRB}Y. J. Bao, H. B. Zhuang, S. Q. Shen, and F. C. Zhang,
Phys. Rev. B \textbf{72}, 245323 (2005).

\bibitem {Zhang-08IJMPB}F. C. Zhang and S. Q. Shen, Inter. J. Mod. Phys. B
\textbf{22}, 94 (2008).

\bibitem {Matianxing}T. Ma and Q. Liu, Appl. Phys. Lett. \textbf{89}, 112102 (2006).

\bibitem {Zarea06PRB}M. Zarea and S. E. Ulloa, Phys. Rev. B \textbf{73} 165306 (2006).

\bibitem {Sinova-SSC}J. Sinova, S. Murakami, S. Q. Shen, and M. S. Choi, Solid
State Communication \textbf{138}, 214 (2006).

\bibitem {Schliemann}J. Schliemann, Int. J. Mod. Phys. B. \textbf{20}, 1015 (2006).

\bibitem {SHC-Inoue}J. I. Inoue, G. E. W. Bauer, and L. W. Molenkamp, Phys.
Rev. B. \textbf{67}, 033104 (2003), \textit{ibid}. \textbf{70}, 041303(R)
(2004) .

\bibitem {SHC-Murakami}S. Murakami, Phys. Rev. B. \textbf{69}, 241202(R) (2004).

\bibitem {SHC-Dimitrova}O. V. Dimitrova, Phys. Rev. B. \textbf{71}, 245327 (2005).

\bibitem {SHC-Raimondi}R. Raimondi and P. Schwab, Phys. Rev. B. \textbf{71},
033311 (2005).

\bibitem {SHC-Grimaldi}C. Grimaldi, E. Cappelluti, and F. Marsiglio, Phys.
Rev. B. \textbf{73}, 081303(R) (2006).

\bibitem {Liuchaoxing}B. Zhou, C. X. Liu, and S. Q. Shen, EPL \textbf{79}
47010 (2007).

\bibitem {SdH2}J. Luo, H. Munekata, F. F. Fang, and P. J. Stiles, Phys. Rev. B
\textbf{41}, 7685 (1990).

\bibitem {Du-09exp}R. R. Du, unpublished, private communication.

\bibitem {Bychkov}Y. A. Bychkov, V. I. Mel'nikov, and E. I. Rashba, Sov. Phys.
JETP \textbf{71}, 401 (1990).

\bibitem {Das}B. Das, D. C. Miller, S. Datta, R. Reifenberger, W. P. Hong, P.
K. Bhattacharya, J. Singh, and M. Jaffe, Phys. Rev. B \textbf{39}, 1411 (1989).

\bibitem {Wilde}M. A. Wilde, D. Reuter, C. Heyn, A. D. Wieck, and D. Grundler.
Phys. Rev. B \textbf{79}, 125330 (2009).

\bibitem {Rashba}E. I. Rashba, Fiz. Tverd. Tela (Leningrad) \textbf{2}, 1224
(1960) [Sov. Phys. Solid State \textbf{2}, 1109 (1960)]; Y. A. Bychkov and E.
I. Rashba, J. Phys. C \textbf{17}, 6039 (1984).

\bibitem {Schliemann03PRB}J. Schliemann, J. C. Egues, and D. Loss, Phys. Rev.
B \textbf{67}, 085302 (2003).

\bibitem {Bao}Y. J. Bao and S. Q. Shen, Phys. Rev. B \textbf{76}, 045313 (2007).

\bibitem {Lucignano08PRB}P. Lucignano, R. Raimondi, and A. Tagliacozzo, Phys.
Rev. B \textbf{78} 035336 (2008).

\bibitem {Landau}L. D. Landau and E. M. Lifshitz, \textit{Quantum Mechanics}
(Pergamon, New York, 1981).

\bibitem {Gerhardts}R. R. Gerhardts, Z. Physik. B \textbf{22}, 327 (1975); J.
Gro$\beta$, and R. R. Gerhardts, Phys. Rev. B \textbf{66}, 155321 (2002).

\bibitem {Vasko}F. T. Vasko and O. E. Raichev, \textit{Quantum Kinetic Theory
and Applications} (Springer, New York, 2005).

\bibitem {Rammer}J. Rammer, \textit{Quantum Transport Theory} (Perseus Books,
Massachusetts, 1998).

\bibitem {Sadovskii}M. V. Sadovskii, \textit{Diagrammatics} (World Scientific,
Singapore, 2006).

\bibitem {Bastin}A. Bastin, C. Lewiner, O. Betbeder-Matibet, and P.
Nozieres,\ J. Phys. Chem. Solids \textbf{32}, 1811 (1971).

\bibitem {Streda}P. St\v{r}eda, J. Phys. C \textbf{15}, L717 (1982).

\bibitem {Shijunren}J. Shi, P. Zhang, D. Xiao, and Q. Niu, Phys. Rev. Lett.
\textbf{96}, 076604 (2006).

\bibitem {Ando}T. Ando, A. B. Fowler, and F. Stern, Rev. Mod. Phys.
\textbf{54}, 437 (1982).

\bibitem {Ando2}T. Ando and Y. Uemura, J. Phys. Soc. Japan \textbf{36}, 959
(1974); T. Ando, \textit{ibid.} \textbf{36}, 1521 (1974); \textbf{37}, 622
(1974); \textbf{37}, 1233 (1974)

\bibitem {Girvin}A. M. M. Pruisken, in \textit{The Quantum Hall Effect},
edited by R. E. Prange and S. M. Girvin (Springer, New York, 1990).
\end{thebibliography}
\end{document}